\documentclass[multphys,vecphys]{promult}

\usepackage{makeidx}     
\usepackage{graphicx}    
\usepackage{multicol}    

\makeindex             

\begin{document}

\title*{Recent GRBs observed with the 1.23m CAHA telescope and the status of its upgrade}

\author{Javier Gorosabel\inst{1}\and Petr Kub\'anek\inst{1,2} \and Martin Jel\'{\i}nek\inst{1} \and Alberto J. Castro-Tirado\inst{1} \and Antonio de Ugarte Postigo\inst{3} \and Sebasti\'an Castillo Carri\'on\inst{4} \and Sergey Guziy\inst{1} \and Ronan Cunniffe\inst{1} \and Matilde Fern\'andez\inst{1} \and Nuria Hu\'elamo\inst{5} \and V\'{\i}ctor Terr\'on\inst{1} \and Nicol\'as Morales\inst{1} \and Jos\'e Luis Ortiz\inst{1} \and Stefano Mottola\inst{6} \and  Uri Carsenty\inst{6}}

\titlerunning{GRBs observed with the CAHA 1.23m  and upgrade of its control system} 
\authorrunning{J. Gorosabel, P. Kub\'anek, M. Jel\'inek, A. J. Castro-Tirado et. al.} 


\institute{Instituto de Astrof\'{\i}sica de Andaluc\'{\i}a (IAA-CSIC),  18008 Granada, Spain.
\and Imaging Processing Laboratory (IPL), Universidad de Valencia, Valencia, Spain.
\and INAF/Brera Astronomical Observatory, Via Bianchi 46, 23807, Merate(LC),  Italy.
\and Universidad de M\'alaga, M\'alaga, Spain.
\and LAEX-CAB (INTA-CSIC); LAEFF, PO Box 78, 28691 Villanueva de la Ca\~nada, Madrid, Spain.
\and Institute of Planetary Research, DLR, Berlin, Germany.
}

\maketitle

We report on optical  observations of Gamma-Ray Bursts (GRBs) followed
up by our collaboration with the 1.23m telescope  located at the Calar
Alto observatory.  The  1.23m telescope is  an old facility, currently
undergoing upgrades   to   enable fully  autonomous   response  to GRB
alerts. We discuss the current status of the control system upgrade of
the 1.23m  telescope.      The    upgrade is  being     done    by our
group\footnote{Our   group is    called  ARAE  (Robotic   Astronomy \&
High-Energy Astrophysics) and is based on members of IAA (Instituto de
Astrof\'isica   de   Andaluc\'ia).    Currently the    ARAE  group  is
responsible  to develop   the  BOOTES network  of robotic   telescopes
\cite{Jeli09}.     See   {\tt   http://www.iaa.es/arae/}     and  {\tt
http://www.iaa.es/bootes/index.php}  for more details.}   based on the
Remote  Telescope System,  2nd   Version (RTS2),  which   controls the
available instruments  and interacts with  the EPICS database of Calar
Alto.  Currently the  telescope  can run fully  autonomously  or under
observer supervision using RTS2.   The fast reaction response mode for
GRB reaction (typically with response  times below 3 minutes from  the
GRB onset) still needs some development and testing.  The telescope is
usually   operated  in  legacy interactive    mode,   with periods  of
supervised   autonomous runs  under   RTS2.  We  show the  preliminary
results of several GRBs followed  up with observer intervention during
the testing phase of the 1.23m control software upgrade.

\section{Introduction}
\label{sec:1}

\begin{figure}[t]
\centering
\includegraphics[height=12cm,angle=-90]{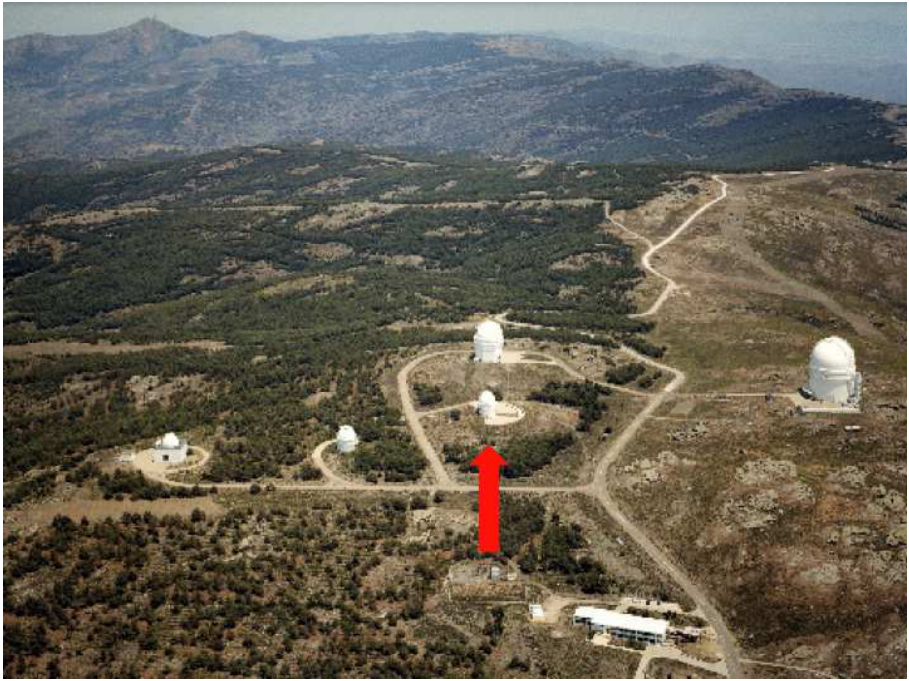}
\caption{ View of the Calar Alto observatory. The picture shows the five telescopes of Calar Alto and the staff building (lower-right corner).  The  arrow shows the  localization  of the 1.23m
telescope. }
\label{CAHA}
\end{figure}

The 1.23m  telescope is at   the German-Spanish observatory of  Calar
Alto (CAHA)  in the province   of Almer\'{\i}a, South-East of  Spain.
The observatory's altitude    (2168   m), mean seeing   (0.9'';   see
\cite{Sanc07}) and a large fraction (65\%) of clear nights, make Calar
Alto one  of  the  most competitive observatories  in  Europe  in the
optical  and near-infrared bands.    The  observatory  harbours  five
telescopes,  but only three are currently  operative: the 1.23m, 2.2m
and 3.5m telescopes.  Fig.~\ref{CAHA} shows  a view of the Calar Alto
observatory. The arrow indicates the position of the 1.23m telescope.

\begin{figure}[t]
\centering
\includegraphics[height=12cm,angle=-90]{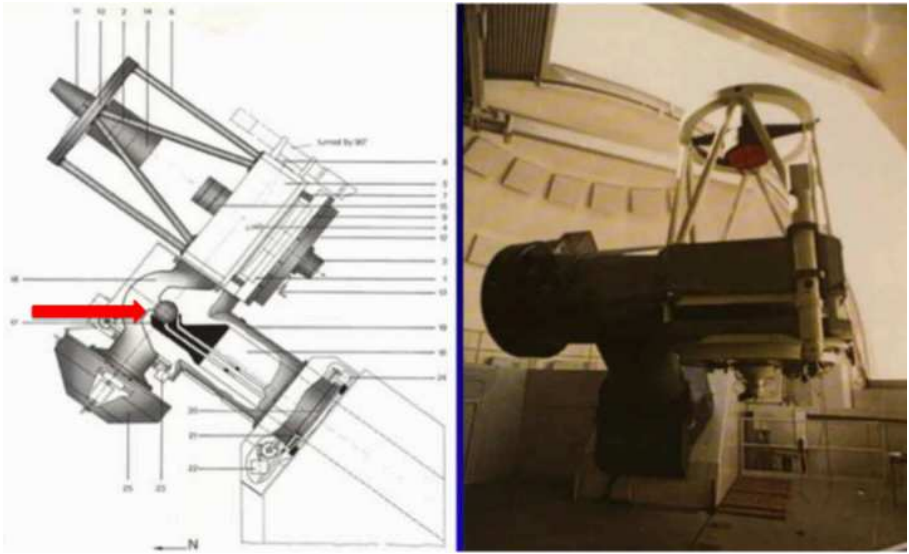}
\caption{ The 1.23m  telescope of  Calar  Alto. 
{\em Left  panel:} the drawing shows the  mechanics and the hydraulic
system  of the 1.23m telescope of  Calar Alto.  One of the mechanical
peculiarities of the  telescope  is the presence  of a  steel  sphere
(indicated by  a  red  arrow)  which, by  means   of a  high-pressure
hydraulic system, supports all  the  weight of  the telescope.   {\em
Right  panel:} Picture  of  the 1.23m  inside the dome.   As seen the
instrumentation is always located  in the Cassegrain focus.  Note the
German mount of the telescope.}
\label{Telescope}
\end{figure}

The 1.23m telescope is a Ritchey-Cr\'etien telescope built in 1975 by
Carl Zeiss.  The  focal ratio of  the telescope  is f/8  with a total
field of view of 90' and a focal plane scale of $\sim 20.9"$/mm.  The
aberration free field  of view is  limited to $\sim 15'$.  The German
mount of the telescope is driven by a mechanical and hydraulic system
renovated  in  2008 by  the  observatory staff.  Fig.~\ref{Telescope}
shows a drawing and a picture of the of 1.23m CAHA telescope.

Two instruments are available  for the 1.23m, the MAGIC near-infrared
camera and  the 2kx2k SITE\#2b optical  CCD.  Currently, most  of the
time ($\sim95\%$) only the optical CCD is used,  as the MAGIC camera,
one of the first near-infrared cameras ever  built, was superseded by
modern instruments on larger telescopes.  In addition to the offering
of     the MAGIC  near-infrared  camera,  CAHA    is considering  the
possibility of installing visitor instruments on the 1.23m telescope.

The field of view of the optical  CCD camera is 17'$\times$17' with a
pixel size of 24$\mu$m. That translates to a  pixel scale of 0.5" per
pixel.  The read-out  time of the whole chip  in 1x1  binning mode is
very long, close to    4 minutes.  For  this reason   rapid follow-up
observations of transients like  Gamma-Ray Bursts (GRBs)  are usually
carried out trimming the window and/or binning the CCD by 2x2 pixels.
Using this technique, the readout time is usually kept below one minute.

\begin{figure}[t]
\centering
\includegraphics[height=12cm,angle=-90]{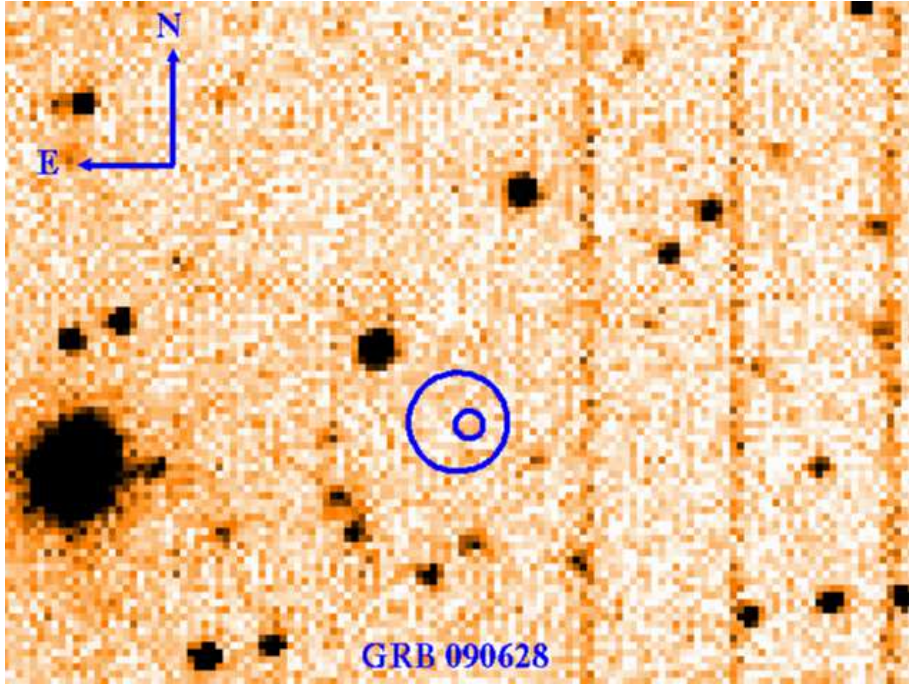}
\caption{Co-added $R$-band image of GRB~090628 taken with the 1.23m telescope.
The large circle represents the preliminary XRT error circle
\cite{Sbar09}, whereas   the   small   one  shows   the   refined  one
\cite{Mang09} reported for GRB 090628.  The field of view of the image
is $125^{\prime\prime} \times 100^{\prime\prime}$.  The total exposure
time is 3120 s.   The individual images  were taken on June 28.9486  -
29.0225  UT (1.43 - 3.20 hours  after the gamma-ray event). No objects
were  found  in both error  circles   down to  $R=22$.  More  detailed
information can be found in \cite{Goro09e}.}
\label{grb090628}
\end{figure}

The CCD  chip is refrigerated  using liquid nitrogen, which makes dark
current negligible,  fewer  than 2  electrons per hour.   The read-out
noise is also low,  about 7 electrons.  However,  the chip has several
bad columns which RTS2 (see next Section and \cite{Kuba06}) must avoid
at the  time of a  GRB alert (see the  vertical lines on  the image of
GRB~090628, Fig.~\ref{grb090628}). The optical camera is equipped with
a $BVRI$ filter wheel.   A Wollaston prism can  be also mounted in the
filter    wheel, which makes  the    1.23m telescope suitable also for
polarimetric studies.

\begin{figure}[t]
\centering
\includegraphics[height=12cm,angle=-90]{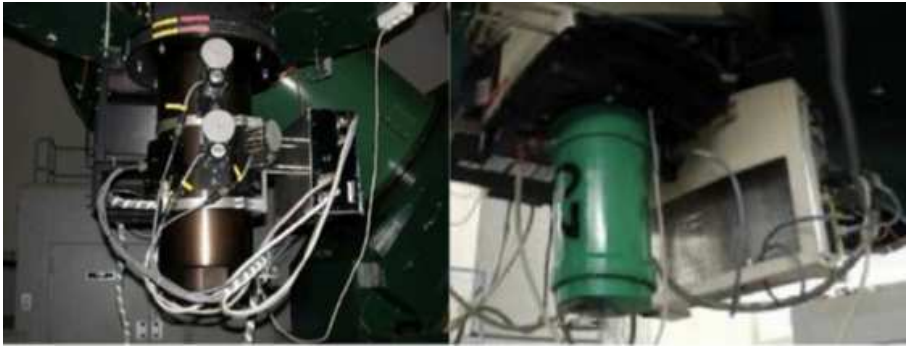}
\caption{The MAGIC near-infrared and optical CCD cameras of the 1.23m
telescope, both  in the Cassegrain  focus of the telescope.  {\em Left
panel:} MAGIC near-infrared  camera.  The MAGIC  detector is a 256x256
pixel HgCdTe  array, providing a field  of view of 4'x4'.   The filter
wheel of MAGIC allows $JHK$ broad-band imaging. The instrument is only
mounted by special request, so is not usually available.  {\em Right
panel:} Optical CCD  camera running  on the  1.23m telescope of  Calar
Alto. The optical  CCD camera is based on  a 2kx2k SITE\#2b chip.  The
green tube is the dewar  and the  white box  is the electronic  system
associated to the  camera.  The filter  wheel is located  in the black
flange just on the detector dewar.}
\label{Instruments} \end{figure}

\section{Upgrading the 1.23m control software with RTS2}
\label{sec:2}

A call for proposals was  issued by CAHA in  2008 for the use of  the
1.23m  telescope over  4 years.  Six  teams obtained  observing time,
having the following scientific drivers:  Solar System bodies, Binary
stars, Transits of exoplanets, T Tauri stars, and GRBs.  The 1.23m is
also  scheduled for public outreach  purposes,  usually associated to
high-schools.

The  ARAE (Robotic Astronomy \&   High-Energy Astrophysics) group  of
IAA\footnote{Instituto  de  Astrof\'isica de Andaluc\'ia,  partner in
the CAHA operations.}  agreed to contribute to  CAHA by providing the
control system of the 1.23m telescope.  The ARAE group was granted 26
GRB triggers  per year  with an average   duration of 0.5 nights  per
trigger.   The upgrade of  the telescope  and  its instrument control
system  is  currently   being  carried    out  based  on    RTS2 (see
\cite{Kuba06} and references therein). One of the key requirements of
the upgrade  is to keep the  existing telescope software and hardware
untouched.  The system must remain operational every night throughout
the upgrade.

As none of the telescope  instruments, nor their control  electronics
and software was developed  by our group, we  are kept away from  the
complex   details of their construction  and  internal operation.  As
described later,  we    exclusively  communicate  with the    control
interfaces of the existing software.

The existing  control  system  is an adapted    version  of that
currently  used by  the 3.5m Calar   Alto telescope.   It  allows the
observer   to control  the     instruments through graphical  user
interface (GUI) programs running  on two main observatory computers  -
one for the telescope control, the other for the camera. The observer
is responsible for preparing the  observing plan, opening and closing
of the dome, taking  care  of executing exposures, inspecting  images
for good pointing and  their quality, and synchronizing the telescope
and the filter  wheel movements with the   CCD exposures.  The  major
drawbacks of this  approach are obvious: the  observer spends most of
the night  working hard to get the  data and keep the system running,
the system is prone to human errors when the observer is  tired.  Moreover, the
observing logs  are either hard to  extract from the technical log or
are not created  by the system  at all, and  the interruption of  the
observation  to react  on  a quickly  evolving target  of opportunity
requires observer presence and attention.

In  contrast, RTS2 was  designed to create  an autonomous observatory
environment. The observer is allowed to interact with the system, and
at worst case  to take  full manual  control.  The system is  able to
guide the  observatory through the night,  taking care of closing and
opening  the dome,  acquiring sky   flats,  darks, and  last, but not
least, keeping detailed logs of the images acquired, judging pointing
accuracy and producing preliminary results.

RTS2 device drivers are responsible for handling any errors that occur
during their operation. If possible, the  device is reset, and another
attempt to  get  it operating  is made.   The  RTS2 {\it rts2-xmlrpcd}
component is responsible for communicating the errors to the users and
custom scripts,  which allows for  the  execution of more  complicated
scenarios.  One of the core  principles is to avoid restarting drivers
that have failed.  If the driver produces  a core dump, that is safely
removed from  the system  and it  is up to    the observer to  restart
it. This feature also allows RTS2 to be  quite flexible -- for example
a configuration without any  telescope, just with  a camera and  other
instrumentation, can  be   made  without changing   a  single  line of
code. More information on handling the errors and other related issues
is provided in \cite{Kubanek09}.

The autonomous capabilities  of the RTS2 system  resides in a generic
layer, with  underlying hardware-specific  drivers and  communication
via the TCP/IP network stack. So in an ideal world, once the provided
skeleton drivers are used  to write low   level RTS2 drivers for  the
hardware, every  telescope can   be  made fully autonomous.   To  our
knowledge,  this is a   big    step forward  from the    traditional,
incremental    way of how   observatory   control  software has  been
developed.  Instead of being  written primarily as  a set of controls
for  the hardware, with  some subsequent  autopilot features, RTS2 was
designed  from    the  start  to  provide    autonomous capabilities.
Secondarily  RTS2 also provides a  way  for the  observer to interact
directly with the hardware.

\subsection{Current state of the upgrade}

Development of the first version of the RTS2 drivers was  a question of a
few days (and  nights), since the  RTS2 drivers were being written by
the main author of RTS2 (with the assistance of the CAHA staff).  The
major obstacle which  we had  to face   was  running RTS2 on an   old
Solaris operating  system,  which is used  to run  the  1.23m control
computers.  As RTS2  was written in quite portable  C++ on Linux, and
using    GNU    Autotools\footnote{GNU   Autotools web     site, {\tt
http://www.gnu.org/software/autoconf}} for build control, the porting
process  involved changing a  few  unavailable functions\footnote{and
renaming the variables  called  "sun" ("sun"  is a defined  symbol on
Solaris  systems).}.  The system is  now able to  operate the same as
any other observatory using RTS2.

The remaining obstacle is the lack  of a natural incorporation of the
auto-guider in  the RTS2  environment.   This fact  prevents  us from
taking images  with long  exposure times.   Tests performed with  the
1.23m  showed that  exposures    longer than 300s   produce elongated
Point-Spread-Functions (PSFs), especially under sub-arc-second seeing
conditions.   The   guider   is  an old    instrument,  with  a quite
complicated interface,  without any autonomous  capabilities.   Thus,
some time  will   be  needed  before  we will   be able   to  perform
observations with the guider smoothly integrated in RTS2.

\begin{figure}[t]
\centering
\includegraphics[height=12cm,angle=-90]{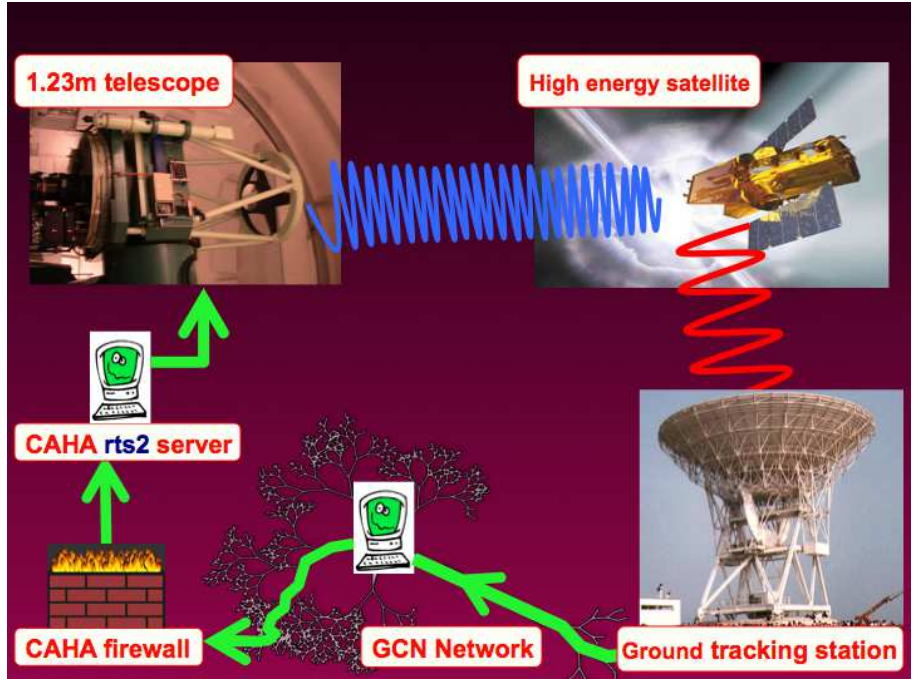}
\caption{Flow diagram of the GCN network and the response mode of the 
1.23m telescope. The whole process can take up to 4 minutes, depending
on  the position of the  GRB on sky.  The  time needed  to receive the
coordinates from  the high-energy satellite in  the RTS2  server takes
just a few seconds  (usually less than 15  seconds). Currently most of
the delay is due  to the slow pointing  of the 1.23m, which could take
up to 4 minutes in the worst case.}
\label{GCN}
\end{figure}

Currently the 1.23m is able to respond to GRB alerts generated by the
GRB Coordinates  Network (GCN).  The maximum  slew time of  the 1.23m
telescope is 4  minutes for the  most unfavourable move.  Usually the
response time is below 3 minutes. The response time is limited by the
speed of  the   current engines/mechanics moving  the   dome and  the
telescope.  Given that the  existing telescope/dome mechanical  parts
are strong and  reliable, CAHA does not plan  to renew them, so we do
expect to overcome the slew-time limitation in  the near future.  The
RTS2   Gamma-Ray  Burst Daemon   ({\it rts2-grbd})  of  the 1.23m  is
continuously linked  via TCP/IP network socket to   the GCN server at
Goddard Space Flight Center (GSFC).  In order to react to GCN alerts,
it  was  necessary to accommodate    the GCN connection  in the  CAHA
firewall.    Fig.~\ref{GCN} shows  a  working   scheme of  the  1.23m
response mode to GCN alerts under RTS2.

When a high-energy satellite (usually the {\it Swift} mission with
its BAT\cite{Bat} detector) localizes a GRB, the position is dumped
in a few seconds from the GRB occurrence to ground-tracking stations
and distributed by the GCN to its subscribers.  In our case the GCN
packet reaches the CAHA {\it rts2-grbd} server.  Then the ongoing
observations are interrupted and the 1.23m is pointed towards the GRB
position in order to start a series of short exposures (usually with
the CCD windowed and binned in order to save read-out time).
Interrupting an observing run is a standard mechanism in RTS2 (see
details in \cite{Kubanek09}).  The interruption can either be hard,
which interrupts the current exposure, or soft, which waits the
current exposure to finish.  The rules governing this choice will be
decided by the observers once the system is fully operational.

The RTS2 control  system stores most  of  the data in  a $Postgresql$
database\footnote{Postgresql           web          site,        {\tt
http://www.postgresql.org}}.   It includes  tools  for retrieving and
manipulating   information from the   database.  New  targets can  be
entered   through  command line   tools,   autonomously from  various
triggering systems, or from the RTS2 Web interface.

The system has a  powerful internal  scripting engine, which  enables
users to define the observing  strategy.  By tracking devices states,
the system handles synchronization among   instruments, so the   user
does not have to know the details of  the execution of the underlying
observation.   Scripting  enables the   user   to specify  all  image
parameters, dithering strategy, selected  filters, and much more. For
example the following script runs  observations in $I$-band, based on
a series of 9 dithered images with a exposure time (per frame) of 240
seconds, with a 1x1 CCD binning and  trimming the CCD to a sub-window
of 500x400 pixels [x1=600, x2=1100, y1=300, y2=700]:
\\
\newline
{\it FW0.filter=I~READT=(600,500,300,400)~binning=1~E~240
\newline
T0.WOFFS=(-72s,+36s)
\newline
for~3~\{ for~3~\{~E~240~\ T0.woffs+=(16s,-54s)~\}~T0.woffs+=(0,+3m) \} }
\newline

Note that the  syntax for the sub-window in the  above example is (x1,
x2-x1,  y1,  y2-y1).  We  have  incorporated in  RTS2  a  call to  the
automatic    astrometric     calibration    package    described    in
\cite{deUgarte}.  This ''on-the-fly'' astrometric calibration corrects
the telescope  pointing, so  the object can  be placed on  the desired
(x,y) coordinates of the detector.   The optical distortion on the CCD
is negligible (less than one tenth  of a pixel), so we decided to make
a simple (and hence fast) fit considering a rotation term and a center
field  shift.  The  astrometric calibration  is done  on  every frame.
Polynomial  fit is  not used.   The typical  astrometric  error (using
$\sim$ 20 stars) is below $0.5^{\prime\prime}$, good enough to correct
the 1.23m  pointing.  The typical  standard deviation in  the inferred
rotation angle  is $\sim0.1$  degrees when $\sim  20$ field  stars are
used in the  fit.  The success rate of  the ``on-the-fly'' calibration
is above 95\%.  The astrometric calibration of the remaining images is
done manually afterwards.

In  the near   future we  plan  to provide  also a  rough photometric
calibration     based  on  the    USNO-B   catalogue   \cite{Mone03}.
Simultaneously to the telescope  triggering an e-mail  is sent to the
1.23m users to  warn them on the  occurrence of the GRB. Additionally
the GCN  alert triggers the creation  of an e-mail and  a SMS for the
ARAE group   with  relevant information    of the GRB   (coordinates,
Galactic  reddening,   finding chart,   elevation  curves, and  other
information). Please see \cite{Castillo}  for more details about this
system.

Further improvements  on this status  are  likely doable, so  we feel
confident that, under  the  current  limitations (for  instance   the
telescope slew speed), we could reduce the  GRB response times.  This
might allow us  to detect the  prompt optical  emission associated to
the gamma-ray event.

Another   potential upgrade  of the   1.23m   telescope could  be the
incorporation  of MAGIC in  RTS2.    Rapid response observations  with
MAGIC would  make the 1.23m  telescope  very competitive  in the  GRB
field.  Currently this is beyond our scopes (and also beyond the CAHA
man-power maintenance  capabilities), but we do  not discard it since
the flexibility of RTS2 to accommodate  new devices would allow us to
integrate MAGIC (and new possible visitor instruments) quickly.

\subsection{Integration of the legacy interfaces to RTS2}

The  following   subsections   provide  descriptions  of   the  legacy
interfaces and their interaction with RTS2.

\subsubsection{Interaction of RTS2 with the Calar Alto EPICS}

The individual operations of  the CAHA telescopes are coordinated, and
controlled, by the  Experimental Physics and Industrial Control System
(EPICS\footnote{EPICS               web            site,          {\tt
http://www.aps.anl.gov/epics/index.php}}).    The task of  EPICS is to
provide access and control of all the CAHA telescopes, as well as data
from the  central weather station and from  the  seeing and extinction
monitors.   The  seeing and  extinction  monitors are   based on small
aperture telescopes located in CAHA  whose values are available through
EPICS.

The 1.23m telescope mount, its focuser and the camera filter wheel are
all fully  controllable through EPICS.  The legacy telescope interface
accepts objects   coordinates in  the  J2000  coordinate  system,  and
handles    all  the   required   calculations   internally,  including
precession,  aberration,  reflection  and telescope   pointing   model
offsets.  Both  the filter  wheel  and the focuser  can be  controlled
through their  own   EPICS channels.    Other devices  in  the   1.23m
telescope system do not exist in the EPICS universe: the CCD detector,
the auto-guider camera, the dome itself, and various auxiliary switches
(for example the dome lights).

We  use the information  provided   through EPICS to automate  several
tasks  of the  1.23m.   For instance,  the meteorological  information
available from the EPICS is used to trigger the bad weather state, and
hence to  close the   dome\footnote{Based on calculated   Sun position
through    {\it libnova},   {\tt   http://libnova.sf.net}}  and   stop
observations.    Bad  weather  is also     triggered  when both  other
telescopes are closed   -  this is   the  rule imposed on    the 1.23m
operations by CAHA.  For discussion on how the weather state voting is
integrated      into       RTS2,        including   its      fail-safe
capabilities\footnote{Network  crashes   and other     failures
(including the   EPICS ones)  are  properly handled  inside  RTS2.   A
detailed description on how the weather  voting system works is beyond
the scope  of this paper.    We  refer the   reader to  the   extended
discussion given in \cite{Kubanek09}.}, please see \cite{Kubanek09}.

Yet another  possible  application  lies in coordinating  observations
with the other Calar Alto telescopes.  From the EPICS system, RTS2 can
learn what the targets  of the other  telescopes are, whether they are
in the  RTS2  database, and depending  on  the target information  can
either start their monitoring or remove them  from the list of targets
which should be observed.

\subsubsection{CCD integration within RTS2}

The optical CCD detector is connected to its own control computer. The
control computer communicates with the  control software, running on a
master  workstation, over the network.  From the available C source code,
we  created a    RTS2 device  driver.  All  the    major  settings are
supported,  including  binning  and partial  chip  readout. The camera
behaves  as   just  another RTS2  supported   CCD  camera, visible  in
monitoring software and available for scripting.

\subsubsection{Guider camera and its integration in RTS2}

The guider is built from a video camera with an image intensifier, fed
from a pick-off mirror on a two-axis  stage.  This allows to place the
guider image  anywhere in  the  telescope  field of view  (FoV),  thus
eliminating the significant disadvantages of autoguiding-by-astrometry
using  the   main camera:  limited  detector FoV,     low gain  and  a
requirement for  short exposure times.  The guider  camera has its own
control computer, which assumes the observer is sitting in front of it
(e.g. the  video output  is  sent directly   to the screen).   Partial
remote control  has  since been provided,  in  part by placing   a web
camera in front of the guiding screen.

We  would like  to improve this   setup  with a fully integrated  RTS2
device, which would  transparently provide automatic search for bright
stars in the  FoV, and guiding  capabilities.   In principle, RTS2  is
able to do this, and some promising tests were  already carried out on
the other RTS2  controlled telescopes.  Currently the biggest  problem
is to figure out how to communicate with the  guider, and to implement
a full auto-guiding loop.

\subsubsection{Dome control and other switches}

As the dome is a critical component,  its control is separate from the
EPICS system (although dome status is reported to EPICS).  In order to
change  the  dome state, special  commands must   be  run on  the dome
control computer. Similar  commands are available  to turn off and  on
dome  lights, to control   the telescope  drives, the  hydraulics, the
tracking and the mirror cover.  Those commands are fully interfaced in
RTS2, so RTS2 is able to control all those switches.

\section{Preliminary results}
\label{sec:3}

 Although not fully   autonomous,  the 1.23m has  already   performed
 follow  up optical observations of GRBs.   None of the results below
 itemized were acquired  by the automatic  response mode  of the RTS2
 package, but  some data were manually acquired  by using RTS2 as the
 observing tool.   Most  of  the data  were  taken by  {\it in  situ}
 observes, using the currently available GUI.

All the   below listed   GRBs showed   X-ray afterglows  which   were
localized by the   XRT  X-ray telescope   on board  {\it Swift}  (see
\cite{xrt} for detailed information on the XRT instrument).  So their
X-ray afterglows were  localized  with  uncertainties of only  a  few
arc-seconds,  making the   corresponding  optical  studies  much  more
efficient.

\paragraph{GRB 090313:} 

The optical afterglow of this GRB \cite{Chor09} was observed with the
1.23m  in the $R$-band during two   consecutive nights one week after
the gamma-ray event.  The observations were accompanied with $K$-band
observations carried  out   with the  3.5m  telescope   of Calar Alto
equipped with   Omega$_{2000}$.    The data   of the  afterglow   are
currently being analyzed and  are part of an international monitoring
which will be published in \cite{Mela09}.

\paragraph{GRB 090424:}

\begin{figure}[t]
\centering
\includegraphics[height=12cm,angle=-90]{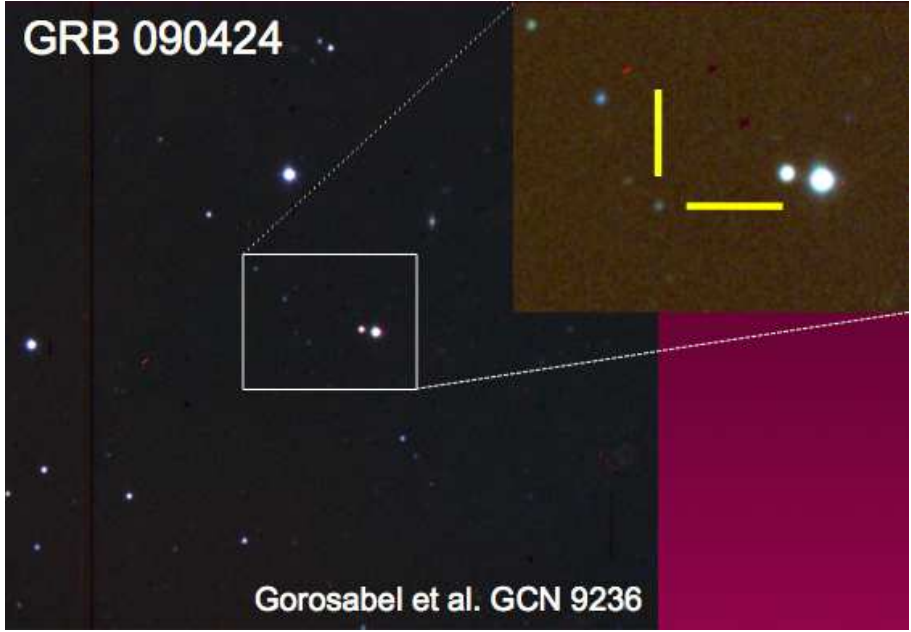}
\caption{The figure shows the optical afterglow of GRB 090424 as detected
with  the  1.23m  CAHA  telescope.   The image   has  been  created by
combining $BVRI$-band images taken with the 2kx2k CCD camera currently
in use. The upper right box shows a magnified picture of the afterglow
region.   The   magnitude of the  afterglow  was   $R=19.3$.  The mean
observing epoch of the image is April 24.87 UT.   North is up and East
right. The field of view of the images is 8'x8'.}
\label{grb090424}
\end{figure}

We detected  the afterglow of GRB 090424  \cite{Cani09}  in $BVRI$ on
April 24.87  UT with a magnitude  of $R=19.3\pm0.1$.  The preliminary
results were  reported in \cite{Goro09a}.  In  the days following GRB
090424, a long term monitoring was performed in $R$  and $I$ bands in
order to search  for the underlying supernova.   Fig.~\ref{grb090424}
shows a colored image constructed with  the $BVRI$ 1.23m images taken
on  April 24.87  UT.   The   final  results  have  been included   in
\cite{Kann05}.

\paragraph{GRB 090621B:}

\begin{figure}[t]
\centering
\includegraphics[height=12cm,angle=-90]{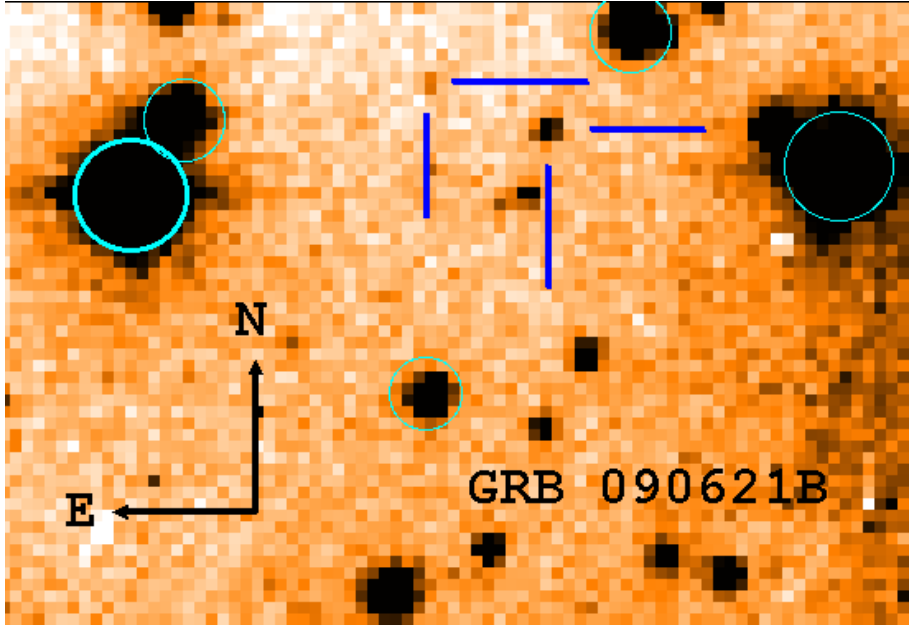}
\caption{The picture shows    the co-added  $I$-band image taken     for
GRB~090621B with the 1.23m telescope of Calar  Alto.  The faintest of
the two objects  (the    most upper one indicated    with tick-marks)
represents the  candidate reported  in \cite{Leva09}.   Approximately
$10^{\prime\prime}$ towards the South-West it is marked the candidate
reported by  \cite{Gale09},  brighter than the  one of \cite{Leva09}.
The  cyan circles  shows the  USNO-A2 catalogue   stars used for  the
astrometric and photometric calibration.   The field  of view of  the
image is  $70^{\prime\prime}\times50^{\prime\prime}$, with North   up
and East left.}
\label{grb090621B}
\end{figure}

We  detected the  two  afterglow  candidates  reported  for this  GRB
\cite{Curr09}   by   Levan  et      al.       and Galeev  et      al.
\cite{Leva09,Gale09}.  The  observations were  done in the  $I$-band,
4.14 - 4.75 hours  after the GRB. A  deep second epoch observation is
pending  in order   to search  for   photometric variability of   the
candidates.    The     preliminary   results   were    reported    in
\cite{Goro09b}. Fig.~\ref{grb090621B} shows both candidates.

\paragraph{GRB 090628:}
$R$-band observations of  the XRT position \cite{Sbar09} were carried
out  1.43--3.20 hours after  the gamma-ray event.  No counterpart was
found  down  to   $R=22$    in  the XRT   error    box.  Simultaneous
near-infrared observations would have  been very helpful in  order to
discriminate    the possible high-redshift   nature  of  this GRB but
unfortunately   they   were  not   possible.   The   results   of the
observations were reported in \cite{Kuba09}.

\paragraph{GRB 090727:}

This  GRB was observed  on  July 27.9646-28.0063  UT in the  $I$-band,
0.45-1.45 hours after the gamma-ray event.  The afterglow reported by
\cite{Smit09} was detected with a magnitude of $I\sim19.4$, using as
calibrator  the   USNO  B1.0  star   with $I=18.72$ and   coordinates
RA$_{J2000}$ = 21:03:49.099, DEC$_{J2000}$ = +64:55:58.23. The results
can be found in \cite{Goro09c}.

\paragraph{The discovery of the optical afterglow of GRB 090813:}

GRB 090813 represents the first optical  afterglow discovered with the
1.23m  CAHA telescope. $I$-band observations of  the XRT position were
initiated 437s after the GRB trigger, revealing an object with a rough
magnitude of $I\sim$17 coincident  with the XRT  position. The lack of
the  object on  the DSS  strongly suggested its   association with the
optical   afterglow of GRB  090813.    The  preliminary  results  were
reported in \cite{Goro09d}.

\paragraph{GRB 090814B:}

This is the only GRB detected by  the INTEGRAL satellite to date that we have
followed up with the 1.23m telescope.  We carried out a series
of $I$-band  observations with different  exposure times  ranging from
180s to 700s. The total exposure time invested for this GRB was 9520s,
with a mean observing epoch of Aug 14.1259 UT.   No optical object was
found  within the refined  X-ray error circle  provided hours later by
XRT.    The 3$\sigma$  limiting magnitude   of the  co-added image  is
$I=20.6$. A more extended description can be found in \cite{Goro09e}.

\section{Conclusions}

The 1.23m  is a wide-purpose telescope  which is currently used by six
international teams to perform  long-term projects with a  duration of
four years.  The ARAE  group of IAA  is responsible for automating the
telescope  operations so that  such  teams can perform their  (usually
long) observing campaigns  without errors, yet enabling quick override
observations of GRBs.

The GRB   results obtained to  date  have been  mostly taken by night
observers.  Use of  the fully autonomous mode,  provided  by RTS2, is
pending  non-trivial integration of the  guider.  After this is done,
we have reasonable  hopes to believe that  telescope will be  able to
react to GRB alerts in a few minutes.  This  could allow us to detect
the optical emission at the first stages of the explosion, making the
associated GRB science much more attractive for the GRB community.

We acquired images  for seven GRBs,  detecting the optical afterglows
of  four of  them.  The typical  reaction  time of these observations
ranged    from $\sim8$ minutes   up   to  a  few  hours.  When  fully
implemented, the  autonomous system   should   be able to   react  to
triggers within 4 minutes.   We have reasonable  hopes to obtain much
more interesting and world--competitive results in the near future.

\vspace{0.5cm}

{\it Acknowledgments}

\vspace{0.2cm}

The research  of JG,  AJCT, RC and   MJ  is supported  by  the Spanish
programmes AYA2008-03467/ESP, AYA2009-14000-C03-01 and  AYA2007-06377.
We are very  grateful to all the CAHA  staff and in particular to Ulli
Thiele for  his excellent support  with the 1.23m telescope.  PK would
like to acknowledge   generous financial support  provided  by Spanish
{\it Programa de  Ayudas FPI del Ministerio  de Ciencia e Innovaci\'on
(Subprograma FPI-MICINN)} and  European {\it Fondo Social Europeo}. We
also would like to thank the  two anonymous referees for their helpful
comments.

\vfill
\eject

\end{document}